\documentclass{hep99}
\input epsf
\begin{document}

\title{The inclusive $B\to \eta' X_s$ decay and $b\to sg^*$
form factors}

\author{Guey-Lin Lin$^{\dag}$}

\address{ Institute of Physics, National Chiao-Tung
University, Hsin-chu 300, Taiwan\\[3pt]
E-mail: {\tt glin@cc.nctu.edu.tw}}

\abstract{We compute the branching ratio of
inclusive $B \to \eta' X_s$ decay based upon the 
QCD anomaly mechanism: $b\to s+g^*\to s+g+\eta'$.
To obtain a reliable $B \to \eta' X_s$ branching ratio, 
we calculate the $b\to s+g^*$
form factors up to the next-to-leading-logarithmic(NLL) approximation.    
We point out that the Standard Model
prediction for $B\to \eta' X_s$ is consistent with the CLEO data, in contrast 
to the claims of some previous works. 
 } 

\maketitle
\fntext{\dag}{based upon a work done with X.-G. He}
% Text of footnotes comes after \maketitle
%\fntext{1}{E-mail: tony.cox@ioppublishing.co.uk}
%\fntext{2}{E-mail: jim.revill@ioppublishing.co.uk}
%\fntext{\dag}{Here is a footnote.}

\section{Introduction}
The observations of exclusive $B\to \eta' K$\cite{CLEO1} and inclusive  
$B\to \eta^{\prime}X_s$\cite{CLEO2} decays with high momentum
$\eta^{\prime}$ mesons have stimulated many theoretical activities
\cite{AS,HT,he1,KP,FR,excl,excl1,HZ}. 
The experimental fitting\cite{CLEO2} shows that
the dominant mechanism for the inclusive mode should be
$b\to sg^*\to
sg\eta^{\prime}$\cite{AS,HT} where the $\eta^{\prime}$ meson is  
produced 
via the anomalous 
$\eta'-g-g$ coupling. According to a previous analysis\cite{HT},
this mechanism within the Standard Model(SM) can only account for 1/3 
of the 
measured branching ratio: ${\cal B}(B\to \eta^{\prime}X_s) =
\left[6.2\pm 1.6(\rm stat)\pm 1.3(\rm syst)^{+0.0}_{-1.5}(\rm bkg)
\right]\times 10^{-4}$\cite{CLEO2} with $2.0 < p_{\eta'} < 2.7$ GeV. 
Furthermore, the subleading mechanism for $B \to \eta' X_s$, 
based upon four-quark operators of the effective  
weak Hamiltonian\cite{he1,KP}, is not sufficient to make up the 
above deficiency. Due to results of these analyses, 
the possibility of an enhanced $b\to sg$ or other 
mechanisms arising from physics 
beyond the Standard Model are raised to account for 
the above discrepancy in branching ratios\cite{HT,KP,FR}. 
In order to see if new physics should play any role in $B\to \eta' X_s$,
one has to have a better understanding on the SM prediction.
In this talk, we present a careful analysis on 
$B\to \eta' X_s$\cite{XGGL} 
using the next-to-leading order effective Hamiltonian. In section 2, we 
illustrate how to compute the off-shell $b\to sg^*$ 
form factors in such a framework. 
In particular, the QCD equation of motion is applied to transform the
charge-radius form factor of $b\to sg^*$ into the structures of 
certain four-quark operators. Therefore the effective 
weak Hamiltonian is shown applicable for computing 
such a form factor. In section 3,
we calculate the branching ratio and the recoil spectrum of $B\to \eta'
X_s $ decay. The results are found to be consistent 
with the CLEO measurement\cite{CLEO2}. Section 4 is the conclusion.     

\section{QCD equation of motion and $b\to s+g^*$ form factors}
The effective Hamiltonian\cite{REVIEW} relevant to the $B\to \eta'  
X_s$ decay is 
given by:
\begin{equation}
H_{eff}=-{G_F\over  
\sqrt{2}}
V^*_{ts}V_{tb}(\sum_{i=1}^{6}C_i(\mu)O_i(\mu)
+C_8(\mu)O_8(\mu)),\nonumber \\
\label{HAMI}
\end{equation}
with 
\begin{eqnarray}
O_1 &=& (\bar{s}_ic_j)_{V-A}(\bar{c}_jb_i)_{V-A},\nonumber \\
O_2 &=& (\bar{s}_ic_i)_{V-A}(\bar{c}_jb_j)_{V-A},  \nonumber \\
O_{3,5} &=& (\bar{s}_ib_i)_{V-A}\sum_{q}(\bar{q}_jq_j)_{V\mp A},\nonumber \\
O_{4,6} &=& (\bar{s}_ib_j)_{V-A}\sum_{q}(\bar{q}_jq_i)_{V\mp A},  \nonumber \\
O_8 &=& -{g_sm_b\over
4\pi^2}\bar{s}_i\sigma^{\mu\nu}P_RT^a_{ij}
b_jG^a_{\mu\nu},
\label{OPER}
\end{eqnarray}
where $V\pm A \equiv 1\pm\gamma_5$.
Precisely speaking, this effective Hamiltonian can be used to
calculate the off-shell $b\to sg^*$ form factors 
which are expressed as
\begin{eqnarray}
\Gamma_{\mu}^{bsg}=&-&{G_F\over \sqrt{2}}  V_{ts}^*V_{tb}
{g_s\over 4\pi^2} (F_1 \bar s( q^2 \gamma_\mu - q\!\!\!/\  
q_\mu) LT^a b\nonumber \\
&-&i F_2 m_b \bar s \sigma_{\mu\nu}q^\nu RT^ab).
\label{DECOM}
\end{eqnarray}
It is easily seen that $F_2 =-2C_8(\mu)$. 
However, the connection between 
$F_1$ and the effective Hamiltonian $H_{eff}$ is less obvious.  
One may acquire some hints by rearranging the QCD penguin operators:
\begin{eqnarray}
\sum_{i=3}^6 C_iO_i&=&2(C_4+C_6)O_V-2(C_4-C_6)O_A\nonumber \\
&+&(C_3+{C_4\over N_c})O_3+(C_5+{C_6\over N_c})O_5,
\label{GLUE}
\end{eqnarray}
where 
\begin{eqnarray}
O_A&=&\bar{s}\gamma_{\mu}(1-\gamma_5)T^a b \sum_{q}\bar{q}\gamma^{\mu}
\gamma_5T^a q,\nonumber \\
O_V&=&\bar{s}\gamma_{\mu}(1-\gamma_5)T^a b \sum_{q}
\bar{q}\gamma^{\mu}
T^a q.
\end{eqnarray} 
Since the light-quark bilinear in $O_V$ carries the quantum number
of a gluon, one expects\cite{AS} $O_V$ give contributions to
the $b\to sg^*$ form factors. In fact, by applying the QCD equation  
of motion, 
$D_{\nu}G^{\mu\nu}_a=g_s\sum \bar{q}\gamma^{\mu}T^a q$,
we have $O_V=(1/ g_s)\bar{s}\gamma_{\mu}(1-\gamma_5)T^a b D_{\nu}
G^{\mu\nu}_a$. 
In this form, $O_V$ clearly contributes to the charge-radius form
factor $F_1$. Let us denote this part of contribution as 
$F_1^a$.
We have $F_1^a=4\pi/ \alpha_s\cdot 
(C_4(\mu)+C_6(\mu))$. We remark that, at the NLL level, 
$F_1$ should also receive contributions from 
one-loop matrix elements. The dominant
contribution, denoted as 
$F_1^b$, arises from the operator $O_2$ where its charm-quark-pair 
meets to form a gluon. In the NDR scheme, 
we find $F_1^b=4\pi/ \alpha_s\cdot 
(\bar{C}_4(q^2,\mu)+\bar{C}_6(q^2,\mu))$ where $q^2$ is the invariant mass
of the gluon and 
\begin{eqnarray}
&&\bar{C}_4(q^2,\mu)=\bar{C}_6(q^2,\mu)\nonumber \\
&=&{\alpha_s(\mu)\over 8\pi}C_2(\mu)\left({2\over 3}+G(m_c^2,q^2,\mu)\right),
\end{eqnarray}
with\cite{COMM}
\begin{eqnarray}
G(m_c^2,q^2,\mu)&=&
4\int_0^1 x(1-x)dx \nonumber \\
&\times& \log\left (
{m_c^2 - x(1-x)q^2\over \mu^2}\right ).
\end{eqnarray}   
We point out that 
$F_1^b$, the $O_2$ contribution, is not negligible.
For $\mu=5$ GeV, one has $F_1^a=-4.03$ and 
Re($F_1^b$)( the real part of $F_1^b$)
ranging
between $-1.5$ and $-3$ for $0.2 < q^2/m_b^2 < 0.7$. The peak value 
Re$(F_1^b)\equiv -3$ occurs at the charm-pair threshold $q^2=4m_c^2$. For
$q^2 > 4m_c^2$, $F_1^b$ develops an imaginary part whose value is
roughly $2i$. 
Concerning previous results on $F_1$, we note that Ref. 
\cite{AS} intends to compue $F_1$ with effective weak Hamiltonian. However,
only the contribution by $F_1^a$ is considered.
Ref. \cite{HT} took $F_1=-5$
which is a result of an one-loop calculation\cite{HSS}. 
Clearly the $q^2$ dependencies of $F_1$ are also absent.   
As we shall see in the next section, the contribution by $F_1^b$, which 
is not included in previous works, can significantly enhance 
the $B\to \eta'X_s$ branching ratio such that SM prediction is
consistent with the CLEO measurement.

\section{The inclusive $B\to \eta' X_s$ decay}
In this decay, the $\eta'$ final state arises from the 
off-shell gluon splitting, $g^*\to g\eta'$, where $g^*$ is 
produced through $b\to sg^*$. The branching-ratio  distribution of
$b(p)\to
s(p')+g(k)+\eta^{\prime}(k^{\prime})$ is\cite{HT}:
\begin{eqnarray} 
{d^2{\cal B}(b\to sg\eta')\over dx dy}
&\cong& 0.2\cos^2\theta
\left( {g_s(\mu)\over 4\pi^2} \right)^2 {a_g^2(\mu)m_b^2\over 4}\nonumber \\
&\times& \left( {\vert \Delta F_1\vert }^2 c_0 + {\rm Re}(\Delta F_1 F_2^*)
{c_1\over y} \right.\nonumber \\
&+& \left.{\vert \Delta F_2\vert }^2 {c_2\over y^2} \right),
\label{BR}
\end{eqnarray}
where $a_g(\mu)\equiv \sqrt{N_F}\alpha_s(\mu)/ \pi f_{\eta'}$ is 
the strength of $\eta'-g-g$ vertex: $a_g\cos\theta
\epsilon_{\mu\nu\alpha\beta}q^{\alpha}k^{\beta}$ with $q$ and $k$ the 
momenta of two gluons; 
$x\equiv (p^{\prime}+k)^2 / m_b^2$ and $y\equiv (k+k^{\prime})^2/  
m_b^2$; 
$c_0$, $c_1$ and $c_2$ are functions of $x$ and $y$  
given by:
\begin{eqnarray}
c_0 &=& \left[ -x^2y+(1-y)(y-x^{\prime})(x+{y\over 2}-{x^{\prime}\over 2})
\right ], \nonumber \\ 
c_2 &=& \left[ x^2y^2-(1-y)(y-x^{\prime})(xy-{y\over 2}+{x^{\prime}\over 2})
\right],\nonumber \\ 
c_1 &=&(1-y)(y-x^{\prime})^2,
\end{eqnarray} 
with $x^{\prime}\equiv m_{\eta^{\prime}}^2 / m_b^2$; and the 
$\eta^{\prime}-\eta$ mixing angle
$\theta$ is taken to be $-15.4^o$\cite{FK}. Following the approach in
Ref. \cite{HT}, we evaluate the $\alpha_s(\mu)$ in $a_g$ at the running 
scale $\mu^2=q^2$. Taking $\mu= 5$ GeV for evaluating 
other scale-dependent quantities,
we find ${\cal B}(b\to sg\eta')=6.4\times  
10^{-4}$
with the cut $m_{X}\equiv \sqrt{(k+p')^2}\leq 2.35$ GeV imposed in 
the CLEO measurement\cite{CLEO2}. This branching ratio is 
consistent with CLEO's measurement on ${\cal B}( B\to \eta' X_s)$
branching ratio\cite{CLEO2}.  
Without the kinematic cut, we obtain ${\cal B}(b\to sg\eta')=1.2\times  
10^{-3}$, which is much larger than $4.3\times 10^{-4}$ 
calculated previously\cite{HT}. 
Clearly this enhancement is due to the contribution of $F_1^b$, 
which increases the magnitude of $F_1$ and thus enhances the 
the branching ratio of $b\to sg\eta'$ according to Eq. (\ref{BR}).   
Since Ref.\cite{AS} also neglects the contribution by $F_1^b$,
its prediction on ${\cal B}(b\to sg\eta')$ would be
much smaller than ours if the 
$\eta'-g-g$ coupling there 
is also evaluated at the running scale $\mu^2=q^2$.
However, the prediction by Ref. \cite{AS} is 
comparable to ours, since, in that work, the
$\alpha_s(\mu)$ in $a_g$ is evaluated at the lowest 
possible scale $\mu^2=m_{\eta'}^2$, and the interference between the
contributions by $F_1$
and $F_2$ is constructive rather than destructive reported here and in 
Ref.\cite{HT}.   

To ascertain our calculation, we also check the $\mu$ dependence of 
the $b\to sg\eta'$ branching ratio. Using NDR scheme with $\mu=2.5$ GeV
and imposing the kinematic cut $m_{X}\leq 2.35$ GeV,
we find ${\cal B}(b\to sg\eta')=7.1\times 10^{-4}$, which is only $10\%$
larger than the branching ratio obtained at $\mu=5$ GeV. This insensitivity
to the scale-choice is reassuring. 
We also obtain the spectrum         
$d{\cal B}(b\to sg\eta')/dm_X$ which has been shown in Ref.\cite{XGGL} and
will not be displayed here.
The peak of the 
spectrum corresponds to $m_X\approx 2.4$ GeV. As pointed 
out in Ref. \cite{CLEO2}, this type of spectrum gives the best fit to the 
CLEO data.
%\begin{figure} [hbt] 
%\begin{center}
%\input{epsf}
%\epsfxsize=6.8cm 
%\epsfysize=5.4cm
%\centerline{\epsfbox{spectrum.eps}}
%\caption{The distribution of ${\cal B}(b\to sg\eta')$
%as a function of the recoil mass 
%$m_X$.}
%\label{fig1}
%\end{center}
%\end{figure}

\section{Concluding remarks}

We have calculated the branching ratio of 
$b\to sg\eta^{\prime}$ by including the NLL correction to the $b\to  
sg^*$ form factors. By evaluating the $\eta'-g-g$ coupling at the running 
scale $\mu=q^2$ and  
cutting the recoil-mass $m_X$ at $2.35$ GeV, we  
obtained
${\cal B}(b\to sg\eta^{\prime})=(6.4-7.1)\times 10^{-4}$ depending on the
choice of the renormalization scale for evaluating the $b\to sg^*$ 
form factors. We have not addressed the possible form-factor suppression of
the $\eta'-g-g$ vertex, which occurs as the gluons attached to the vertex
go farther off-shell\cite
{AS,HT,KP}. So far
it remains unclear how much  
the form-factor suppression might be.
However, comparing our prediction with the CLEO measurement on ${\cal B}
(B\to \eta' X_s)$, which still has a large error bar,
it remains possible  
that the anomaly-induced 
process $b\to sg\eta^{\prime}$ could account for the CLEO data within the 
framework of the Standard Model.    

\section{Acknowledgments}  
This work is supported by
National Science Council of R.O.C. under the grant number 
NSC 87-2112-M-009-038,  
and National Center for Theoretcal Sciences of  
R.O.C. 
under the topical
program: PQCD, B and CP.


\begin{thebibliography}{11}
\bibitem{COMM}
In the original article\cite{XGGL}, we use the scheme-independent 
Wilson coefficient\cite{BJLW}. 
Hence the contribution of one-loop matrix element 
is also distinct. In particular, the constant added to $G(m_c^2,q^2,\mu)$
in Eq. (6) is ${10\over 9}$ rather than ${2\over 3}$. For detail, we refer the
readers to Eqs. (4) and (5) of Ref.\cite{XGGL}. Note that the 
r.h.s. of Eq. (4) in Ref.\cite{XGGL}
should be $P_s(q^2,\mu)$. The extra minus sign is a typing 
error.
\end{thebibliography}
\end{document}